\documentclass[review]{elsarticle}

\usepackage{xspace}
\usepackage{graphicx}
\usepackage{float}
\usepackage[utf8]{inputenc}
\usepackage{hyperref}

\usepackage{array}
\newcolumntype{P}[1]{>{\centering\arraybackslash}p{#1}}

\newcommand{\strech}{-2.5pt}

\newcommand{\discr}{\mbox{DISCR}\xspace}

\newcommand{\descr}{\mbox{DESCR}\xspace}

\newcommand{\fddscore}{\mbox{FDD$_{\beta}$}\xspace}

\newcommand{\precision}{\mbox{precision}\xspace}

\newcommand{\recall}{\mbox{recall}\xspace}

\newcommand{\fscore}{\mbox{F$_{\beta}$}\xspace}

\newcommand{\fonescore}{\mbox{F$_{1}$}\xspace}

\newcommand{\MAX}{\mbox{max}\xspace}

\begin{document}

\begin{frontmatter}
	

\pagestyle{headings}  



\title{Assessing the behavior and performance of a supervised term-weighting technique for topic-based retrieval}

\author[dcic]{Mariano Maisonnave}
\ead{mariano.maisonnave@cs.uns.edu.ar}

\author[imbb]{Fernando Delbianco}
\ead{fernando.delbianco@uns.edu.ar}
\author[imbb]{Fernando Tohmé}
\ead{ftohme@criba.edu.ar}
\author[dcic]{Ana Maguitman}
\ead{agm@cs.uns.edu.ar}

\address[dcic]{Departamento de Ciencias e Ingeniería de la Computación, Universidad Nacional del Sur, Instituto de Ciencias e Ingeniería de la Computación (UNS-CONICET), Bahía Blanca, Argentina}
\address[imbb]{Departamento de Economía, Universidad Nacional del Sur, Instituto de Matematica de Bahía Blanca (UNS-CONICET), Bahía Blanca, Argentina}

\begin{abstract}


This article analyses and evaluates \fddscore, a supervised term-weighting scheme that can be applied for query-term selection in topic-based retrieval. \fddscore weights terms based on two factors representing the descriptive and discriminating power of the terms with respect to the given topic. It then combines these two factor through the use of an adjustable parameter that allows to favor different aspects of retrieval, such as precision, recall or a balance between both.  The article makes the following contributions:  (1) it presents an extensive analysis of the behavior of \fddscore as a function of its adjustable parameter; (2) it compares \fddscore against eighteen traditional and state-of-the-art weighting scheme; (3) it evaluates the performance of  disjunctive queries built by combining terms selected using the analyzed methods;  (4) it introduces a new public data set with news labeled as relevant or irrelevant to the economic domain.  The analysis and evaluations are performed on three data sets: two well-known text data sets, namely {\em 20 Newsgroups} and {\em Reuters-21578}, and the newly released data set.  It is possible to conclude that despite its simplicity, \fddscore is competitive with state-of-the-art methods and has the important advantage of offering flexibility at the moment of adapting to specific task goals. The results also demonstrate that \fddscore offers a useful mechanism to explore different approaches to build complex queries.

\end{abstract}

\end{frontmatter}

\vspace{-0.80cm}	

\section{Introduction}\vspace{\strech}
\label{sec:introduccion}
    
    Topic-based text retrieval is the problem of seeking material stored in the form of text (e.g., documents, news reports, tweets, etc.) relevant to a given topic and returning this material to the user. In order to retrieve topic-relevant results,  topical queries need to be formulated by strategically selecting terms that help achieve good retrieval performance.  Terms can be selected from different sources, depending on how the topic of interest is represented. For instance, a topic can be represented by  a fragment of text, a set of words, or a collection of  documents labeled as relevant or irrelevant to the topic. If such a collection of labeled documents is available, it becomes possible to apply a supervised approach to build topical queries.


    
    In our work we consider broad topics  (e.g. economy, religion, electronics, etc.) represented by collections of documents (e.g., news, newsgroup posts, etc.)  labeled as relevant or irrelevant to the given topic. We aim at extracting different types of variables from the text (terms, entities, events, and others) and to rank them according to their importance to the topic at hand. For instance, a term may be important as a descriptor of the topic or it may have good discriminative power, rendering it as a useful query term.  Our ultimate goals are to provide the user with a set of relevant variables that summarize the given topic and to explore strategies to build topic-based queries based on the selected terms.

    Traditional information retrieval models typically apply unsupervised approaches to determine term importance.  These approaches assign a numerical value to each term in a document based on the number of occurrences of the term in the document. Usually, the higher the frequency of the term in a document the better is its descriptive power. In addition, most models adopt some notion of term specificity, which is usually associated with the number of documents where the term occurs. In this sense, the lower the frequency of a term in a document collection, the higher is its specificity and its discriminating power.  This gives rise to term-weighting  schemes that rely on a combination of term frequency and term specificity  to account for the informativeness of a term in a document.  This informativeness value attempts to represent how well the term allows to describe and discriminate a given document but is independent of any topic of interest. Hence the derived weighting schemes represent the informativeness value of a term for a document as opposed to the informativeness value of a term  for a topic.  This is the case of the widely-used TF-IDF weighting scheme, where terms are weighted based on  local  (TF) and global (IDF) term frequencies. Similar to TF-IDF, other  term-weighting schemes that combine some form of descriptive and discriminating values have been proposed in the literature~\citep{robertson1976relevance,tokunaga1994text,robertson2004understanding,deisy2010novel}. However, no topic labels are used to compute these values.

    Other term-weighting methods take a supervised approach to assess the importance of a term in a topic or class. In most cases, supervised approaches to term weighting have been formulated in the context of classification tasks, where the importance of a term  in a class is a fixed value~\citep{debole2004supervised,lan2005comparative,wang2013inverse,deng2014study,chen2016turning,verberne2016evaluation,fattah2016combined,wu2017balancing,feng2018probabilistic}.  While using a fixed weight may be appropriate for a classification task, it represents a limitation for topic-based query construction since a term may be more or less effective depending on whether the retrieval task requires high recall, high precision or a balance of both.

    In this work, we analyze a supervised term-weighting scheme called \fddscore proposed in~\cite{maisonnave2019flexible} that uses an adjustable parameter $\beta$ to favor different aspects of retrieval.  The \fddscore scheme offers an advantage over several state-of-the-art term-weighting schemes as its parameter allows  to favor either precision or recall, as well as to achieve a balance between both. In addition, the \fddscore scheme relies on a very simple formulation, which is directly derived from basic notions traditionally used  to measure retrieval effectiveness. This is in contrast with most state-of-the-art weighting schemes  that rely on more elaborated information-theoretic or statistical notions, such as entropy, mutual information, information gain and probability distributions. We argue that these and other information-theoretic and statistical notions that are pervasive in the most widely adopted weighting schemes are useful for deriving unsupervised term-weighting methods but not necessarily more effective than the simpler notions adopted in the definition of \fddscore for a supervised scenario.

    The article is structured as follows. Section~\ref{sec:objectives} outlines the objectives and contributions of this work. Section~\ref{sec:background} presents background concepts and reviews related work. Section~\ref{sec:fddscore}  describes the \fddscore measure and discusses how  \fddscore can be applied to define topic-based retrieval strategies from labeled collection of documents  to  favor precision, recall, or a balance of both. 
    Section~\ref{sec:analysis}  examines the effect of the adjustable parameter used in the definition of \fddscore. Section~\ref{sec:evaluation} evaluates  \fddscore  by comparing it with other state-of-the-art weighting schemes, followed by a discussion of the results and their implications in section~\ref{sec:discussion}. 
    Finally, section~\ref{sec:conclusion} presents the conclusions and outlines future research work.
    
    \section{Objectives and contributions}
    \label{sec:objectives}
    
    The overall goal of this research work is to assess the behavior of the \fddscore term-weighting score and its performance for query-term selection in topic-based retrieval. 
Such an assessment requires tackling some important problems, which are the specific objectives of this work:
\begin{enumerate}
\item Understanding the behavior of \fddscore as a function of its tunable parameter $\beta$.  
\item Assessing the effectiveness of \fddscore for query-term selection. The goal is to compare its performance to that of other traditional and state-of-the-art term-weighting methods.   
\item Exploring the usefulness of \fddscore for generating multi-term queries for topic-based retrieval.
\end{enumerate}
 
    
    In previous work, we provided an initial formulation and investigation of the \fddscore measure~\citep{maisonnave2019flexible}. Our preliminary analysis was focused on  the use of  \fddscore for variable extraction in the economic domain. The  \fddscore measure was empirically evaluated through human-subject experiments  both as an estimator of human-subject assessments of term importance and for retrieval effectiveness.  
    While the previous analysis was limited to a single domain (Economy), the  present work offers a major contribution towards the generalization of \fddscore  by performing a deep analysis of the effectiveness of \fddscore for topic-based query construction in  the economic and general domains.

    This article makes the following major contributions: 
    \begin{enumerate}
    
    \item An extensive analysis of the behavior of \fddscore as a function of its adjustable parameter $\beta$.
    
    \item A comparison of  \fddscore against eighteen traditional and state-of-the-art weighting scheme. The methods selected for comparison include both unsupervised and supervised term-weighting schemes.
    
    \item An evaluation and comparison of the performance of  disjunctive queries built by combining terms selected using \fddscore and other state-of-the-art methods.

    \item A publicly available  data set with news  from the {\em The Guardian} newspaper labeled  by domain-experts as relevant and not relevant to the economic domain.
   
    \end{enumerate}
    The analysis and evaluations are performed on three data sets: two well-known text data sets, namely {\em 20 Newsgroups} and {\em Reuters-21578}, and the newly released data set {\em Economic Relevant News from The Guardian}. 
    The results derived from the completed analysis and evaluations demonstrate that the \fddscore weighting scheme offers a useful mechanism to explore different approaches to build complex queries.  Also, the comparative performance evaluation shows that despite its simplicity, \fddscore achieves results that are competitive with state-of-the-art methods that rely on more complex information-theoretic and statistical notions.

\section{Background and related work}	
\label{sec:background}
    
Topic-based retrieval is the process of seeking and retrieving material related to a topic of interest. A collection of labeled documents can be used to represent the topic of interest and used to guide the retrieval process. Topic-based retrieval is different from the most widely known ad-hoc retrieval. Note that the purpose of ad-hoc retrieval is to obtain relevant documents in response to a question~\cite{voorhees1999overview}.  Different from ad-hoc retrieval, the goal of topic-based retrieval is to obtain relevant documents based on a {\em broader topic} specified by a topical context. The topical context can be defined by a sample of relevant documents, which can be augmented by a sample of documents marked as irrelevant to facilitate the training process. This collection can be explicitly provided to the topic retrieval tool or it can be inferred from the long or short standing interests of a user. For instance, a system may be able to automatically identify the topics of interest to a user by monitoring the user interaction with software applications being used (e.g., by tracking  browsing and query history, dwell time,  clicks, etc.). 

Topic-based retrieval can be used to generate topic-based alerts  \cite{eckroth2019genetic}, support expert as they organize domain knowledge \cite{lorenzetti2016mining}, and build vertical portals \cite{pinho2018web}, among many other applications.   
Depending on the application at hand, the emphasis of topic-based retrieval can be placed on attaining high precision, high recall or a combination of both.  Topic-based retrieval usually relies on automatic query formulation. Hence, one of the most challenging problems in topic-based retrieval is to automatically formulate queries where the emphasis can be put on different retrieval goals depending on the given application.

    The problem of topic-based query formulation can be formulated as an optimization problem where the goal is to maximize query effectiveness in terms of specific topic-based retrieval performance metrics.  A difficulty in dealing with query optimization is that this problem does not have optimal substructure, which means that optimal solutions cannot be constructed efficiently from optimal solutions to its subproblems. As a consequence, combining good query terms may not necessarily result in an effective longer query. 
    However, due to the high dimensionality of the search space and to the combinatorial nature of the problem of optimal query formulation,   the most commonly used algorithms for term selection  adopt a na\"{\i}ve approach to the absence of optimal substructure and apply a greedy strategy for query construction. In other words, terms are selected independently of other terms being selected.
    Consequently, query-term selection algorithms typically compute, for each candidate term, a measure of informativeness based on a term-weighting scheme, arrange the terms in decreasing order based on this measure and use some of the best terms to build queries.  The problem of topic-based query formulation then reduces to the problems of defining a term-weighting scheme and deciding how many terms and how these terms will be combined to build a query (e.g., using disjunctive, conjunctive, or a combination of both types of queries).

    Several measures of term informativeness have been used in text classification and information retrieval.  Traditional term-weighting methods take an unsupervised approach and originated in the information retrieval community. According to  \citep{salton1988term}  local and global factor need to be considered to define a term-weighting method, while a normalization factor is sometimes used to correct the weights.  Local factors usually represent whether the term appears in the document an the number of times it does. They are usually designed to improve recall and the rationale behind these factors is that frequent terms are semantically close to the content of the document. The simplest and most commonly used local factors are the binary factor, which only measures the presence (with value 1) or absence (with value 0) of the term in the document, and  the {\em term frequency}  (TF) factor, which counts how many times a term appears in a document. Some variants of  the classical TF scheme are  {\em inverse term frequency} (ITF)~\citep{leopold2002text}, which normalizes the values to the interval [0,1] based on  Zipf's Law,  and other  transformations on the term-frequency values where terms that are extremely frequent do not increase at the same rate as in TF~\citep{debole2004supervised}.  In \cite{dogan2019term}, the authors demonstrate the measurable impact of incorporating local factors in supervised term-weighting schemes. Their analysis include the use of three different local factors (term frequency TF, the square root of term frequency SQRT\_TF and the logarithm of term frequency LOG\_TF) in seven supervised term-weighting schemes. Their experimental results indicate that SQRT\_TF offers promising results as a local factor. 
    	
    Different from local factors, global factors represent how frequent the term is in the document collection, in such a way that frequent terms are penalized. Global factors are designed to improve precision although this might be at the expense of a drop in recall. The rationale behind these factors is that common terms are poor discriminators. This is the case of the widely used global factor {\em inverse document frequency} (IDF)~\citep{salton1988term}, which penalizes terms based on the number of document containing the term. A similar global weighting schemes known as {\em weighted inverse document frequency} (WIDF)~\citep{tokunaga1994text} penalizes frequent terms by taking into account the number of times they occur in each document of a collection.   An unsupervised approach to  approximate the descriptive and discriminative power of a term in a topic is presented in~\citep{maguitman2004dynamic}.

    More closely related to the problem of topic-based term weighting are those  methods that take advantage of class-label information, which give rise to supervised term-weighting schemes.  A simple method that uses class information can be computed by counting the number of documents in a class that contain the term. For instance, TGF* refers to a basic supervised term-weighting method that counts the number of documents in a class that contain the term. A traditional supervised term-weighting score is given by the conditional probability of a term occurring given a class, which gives rise to  the {\em odds ratio} (OR)~\citep{rijsbergen1981selection}. Another probabilistic technique (Prob)~\citep{liu2009imbalanced}  involves two ratios directly related  to the term's strength in representing a category. The first ratio increases with the number of documents of the class  containing that term  while the other is higher when most of the term occurrences are within the class.
    To account for the discriminating power of a term, other supervised schemes such as 
    the {\em inverse class frequency factor} (ICF) and  the {\em category relevance factor} (CRF)~\citep{deng2002linear} penalize a term proportionally to the number of different classes in which the term appears.
    Supervised term-weighting scores derived from information theory include   {\em mutual information} (MI), {\em chi-squared} ($\chi^2$), {\em information gain} (IG) and {\em gain ratio} (GR).  
    
     The {\em Galavotti-Sebastiani-Simi} coefficient (GSS)~\citep{galavotti2000experiments}  and {\em entropy-based category coverage difference} (ECCD)~\citep{largeron2011entropy}  are term-weighting schemes adapted from feature selection techniques. Another scheme uses a {\em relevancy frequency factor} (RF)~\citep{lan2009supervised} to favor terms whose frequency in the positive class is higher than in the negative one.

    Several other definitions of global factors are modifications of the   IDF factor. For example, in \cite{tang2020improved}, the authors propose a new term-weighting technique called term-frequency inverse exponential frequency (TF-IEF), for addressing some limitations found in the classical IDF factor. 
    Another variant of IDF is called  {\em inverse document frequency excluding category} (IDFEC) \citep{domeniconi2015study}. This technique penalizes  frequent terms but avoids penalizing those terms occurring in several documents belonging to the relevant class.  The combination of IDFEC and RF results in the IDFEC\_B scheme~\citep{domeniconi2015study}. 
     In \cite{samant2019improving} the authors propose  ifn-tp-icf, RFR and modOR, which are based on three state-of-the-art techniques, namely iqf-qf-icf~\cite{quan2010term}, RF~\citep{lan2009supervised} and OR~\cite{rijsbergen1981selection}, respectively. They also propose a
      new  method called  ifn-modRF and analyze the performance of the proposed methods in the short text classification task. Other  term-weighting methods specially proposed for the context of short texts include \cite{samant2019improving,alsmadi2019term}.
    In \cite{alshaher2020new}, the authors propose a statistical transformation, namely standardization, that is used to define a new term-weighting technique.
    Another weighting method for computing the discriminating power of a query term presented in~\cite{song2012novel} relies on the ranks or the similarity values of the relevant and non-relevant documents retrieved by a query.

  \emph{Inverse Gravity Moment} (IGM) \cite{chen2016turning} is a term-weighting scheme that incorporates a new statistical model to measure the inter-class distribution concentration of a term. To define IGM for a term $t_k$ it is necessary to compute the term frequency  in each class. The term frequency distribution is used to assess the class distinguishing power of the term, resulting in a ranking of classes for each term. In other words, the frequencies of a term in each class defines a ranking  $f_{k 1} \geq f_{ k 2} \geq \ldots \geq f_{km}$, where $f_{kr}$   ($r$ = 1, 2, $\ldots$, m) is the frequency of term $t_k$ in the $r$-th class  after being sorted.  If the inter-class distribution is uniform, then the center of gravity is located at the center position of the ranking list, and the class distinguishing power is minimal. On the other hand, if the term occurs in only one class, the center of gravity will be at the starting position ($r$ = 1). The position of the gravity center is used to define the IGM metric as follows:
    \[ \mbox{IGM}(t_k) =  \frac{f_{k1} } { \sum_{r=1}^m{f_{kr} \times r}},\]
where $r$ is the rank, $f_{kr}$ is the frequency of term $t_k$ in the $r$-th class after being sorted, and $m$ is the number of classes.  Note that $f_{k1}$ is the frequency of term $t_k$ in the class that has the highest ranking ($r=1$). 
 
The metrics TGF* and IGM can be combined to define TGF*-IGM as follows:
\[\mbox{TGF*-IGM}(t_k) =  \mbox{TGF*}(t_k) \times \left ( 1 +\lambda \times \mbox{IGM}(t_k) \right ) .\] 
 We will set $\lambda=7$  in the evaluation reported later as that is the default value used in \cite{chen2016turning}.  

    In \cite{dogan2019improved} the authors propose a modification of IGM with the purpose of improving its weighting behavior, especially for some extreme cases. The authors pose several scenarios in which the outcome of the IGM measure for different terms that appear in a different number of documents but in the same number of classes is the same. For instance, two terms that appear in only one class will have the same IGM (i.e., $f_{k1}/(f_{k1}\times 1)=1$), regardless of the value of $f_{k1}$. However, one of those terms can occur 100 times in the class while the other may occur only once. In light of these scenarios, the authors incorporate the number of documents in the class in which $t_k$ occurs most ($D_{total}(t_{k\_max})$) to the original IGM formulation, resulting in the following improved formulation of IGM: 
    \[ \mbox{IGM}_{imp}(t_k) =  \frac{f_{k1} } { \sum_{r=1}^m{f_{kr} \times r + \log_{10}{\left ( \frac{D_{total}(t_{k\_max})))}{f_{k1}} \right )}}}. \]
    Then the authors combine IGM$_{imp}$ with the term frequency and the squared root of the term frequency values as follows: 
    \[\mbox{TGF*-IGM}_{imp}(t_k) =  \mbox{TGF}*(t_k) \times \left ( 1 + \lambda \times \mbox {IGM}_{imp}(t_k) \right ),\]
    \[ \mbox {SQRT-TGF*-IGM}_{imp}(t_k) =  \sqrt{\mbox {TGF*}(t_k)} \times \left ( 1 + \lambda \times \mbox {IGM}_{imp}(t_k) \right ) .\] 
In our evaluations we replicate these two weighting techniques using $\lambda=7$ as suggested in \cite{dogan2019improved}.

    Table~\ref{table:definition} shows the  definitions of the main scores presented above. The following notation, adapted from~\citep{lan2005comparative,domeniconi2015study}, is used whenever it is possible:
    \vspace{-0.2cm}
    \begin{itemize}
    	\setlength\itemsep{-0.5em}
    \item $A$ denotes the number of documents that belong to class $c_k$ and contain term $t_i$.
    \item $B$ denotes the number of documents that belong to class $c_k$ but do not contain the term $t_i$.
    \item $C$ denotes the number of documents that do not belong to class  $c_k$ but contain the term $t_i$.
    \item $D$ denotes the number of documents that do not belong to class $c_k$ and do not contain the term $t_i$.
    \item $N$ denotes the total number of documents in the collection (i.e., $N=A+B+C+D$).
    \end{itemize}
    \vspace{-0.2cm}
    \noindent
    Note that some formulations include the expression $\MAX(X,1)$ to prevent the possibility of undefined values, such as divisions by zero or $\log(0)$.

    \begin{table}
    	{\footnotesize
    		\renewcommand{\arraystretch}{0.9}
    	\begin{tabular}{ l l }
    		\hline
    		Scheme & Formulation \\ 
    		\hline		  
    TGF & $ A+C $ \\
    IDF & $ \log{(N/(A+C))} $ \\  
    TGF* & $A $ \\
    TGF*-IDFEC & $A \times (\log((C+D)/\MAX(C,1))) $ \\	
    $\chi^2$ & $ N ((AD-BC)^2/((A+C)(B+D)(A+B)(C+D)))$ \\
    OR & $\log((\MAX(A,1) \times D)/\MAX(B \times C,1)) $\\
    IG &  $ (A / N) \log(\MAX(A,1)/(A+C))- $\\
         & \ \ \ \ \ \ \ \ \ \ \ \ \ \ \ \ \ \ \ $ ((A+B)/N)  \log((A+B)/N) +  (B/N) \log(B/(B+D))$ \\
    GR & $IG/(-((A+B)/N)\log((A+B)/N) -((C+D)/N)\log((C+D)/N))$ \\
    GSS & $\log(2 + (  (A+C+D)   /   (\MAX(C,1))  ))$ \\
    Prob & $ \log(1+(A/B)(A/C)) $ \\
    RF  & $ \log(2+(A/\MAX(C,1)) $ \\	
    IDFEC & $ \log((C+D)/\MAX(C,1)) $ \\
    TGF-IDFEC & $ (A+C)(\log((C+D)/\MAX(C,1)) ) $ \\
    MI & $\log((N \times \MAX(A,1))  /   ((A+B)  (A+C))   )$ \\		
    IDFEC\_B & $ \log(2 +(A+C+D)   /   (\MAX(C,1)) )$ \\
    TGF*-IGM &  TGF*$(t_k) \times \left ( 1 +\lambda \times \mbox{IGM}(t_k) \right )$ \\
    TGF*-IGM$_{imp}$ &  TGF*$(t_k) \times \left ( 1 + \lambda \times \mbox{IGM}_{imp}(t_k) \right )$ \\
    SQRT-TGF*-IGM$_{imp}$ &  $ \sqrt{\mbox{TGF*}(t_k)} \times \left ( 1 + \lambda \times \mbox{IGM}_{imp}(t_k) \right )$\\ 
    
    		\hline
    \end{tabular}
    
    }
    \caption{Definitions of term-weighting schemes.}
    \label{table:definition}
    \end{table}
    
    \vspace{-0.2cm}

\section{The \fddscore term-weighting score }
\label{sec:fddscore}
    
    The \fddscore scheme, analyzed in this article, is a  term-weighting  score that relies on two principles: (1) class or topic labels convey useful information for term weighting, and (2)  the importance of a term depends on the specific objectives at hand (e.g., attaining high recall, high precision or a balance of both).  The result is a parameter-based supervised method that distinguishes two relevancy scores. The first score, to which we refer to as {\em descriptive relevance} (\descr), is local to the class and represents the importance of a term to describe the class. Given a term $t_i$ and a class $c_k$ the \descr score is expressed as:
    \[\descr(t_i,c_k)=\frac{|d_j: t_i \in d_j \wedge d_j \in c_k|}{|d_j:  d_j \in c_k|},\]
    \noindent
    which is equivalent to  $A/(A+B)$, using the notation adopted in the previous section, and stands for the fact that those terms that occur in many documents of a given class are good descriptors of that class.  
    
    The second score represents the {\em discriminative relevance} (\discr).  This score is global to the collection and is computed for a term $t_i$ and a class $c_k$ as follows:
    \[\discr(t_i,c_k)=\frac{|d_j: t_i \in d_j \wedge d_j \in c_k|}{|d_j:  t_i \in d_j|},\]
    \noindent
    which is equivalent to  $A/(A+C)$ and stands for the fact that terms that tend to occur only in documents of that class are good discriminators of that class.

    The \fddscore scheme combines the \descr and \discr scores as follows:
    \[\fddscore(t_i,c_k) = (1+\beta^2)\frac{\discr(t_i,c_k)\times\descr(t_i,c_k)}{(\beta^2\times \discr(t_i,c_k))+\descr(t_i,c_k)},\] 
    \noindent
resulting in the following definition according to the previously adopted notation:
    \[\fddscore(t_i,c_k) = (1+\beta^2)\frac{A/(A+C)\times A/(A+B)}{(\beta^2\times A/(A+C))+A/(A+B)}.\] 
    
    The tunable parameter  $\beta$ offers a means to favor different objectives in the information retrieval task. By using a $\beta$ value higher than 1  we can weight descriptive relevance higher than discriminative relevance while  a $\beta$ smaller than 1 weights discriminative relevance higher than descriptive relevance.
    The \fddscore  is derived from the \fscore formula traditionally used in information retrieval, which measures the effectiveness of retrieval with respect to a user who attaches $\beta$ times as much importance to recall as precision~\citep{rijsbergen1979information}.

    Topic-based retrieval can be formulated as a supervised learning problem where the topic of interest is defined by a training collection of documents labeled as relevant or irrelevant to the topic of interest.  The training collection can be used to compute \fddscore for the terms in the collection as described above. Terms do not need to be limited to single words and may also include  n-grams, concepts or more complex terms.  
    
     The \fddscore score offers a means to identify useful terms based on the training collection and to classify those terms based on their ability to reflect the user's needs. 
    The terms learned from the training collection can then be used to build queries with the purpose of retrieving material from the Web, Twitter or any other unlabeled collection of text resources. Intuitively, if the user is seeking for specific resources then precision can be favored by building queries containing high \fddscore terms for small $\beta$ values. This approach will prioritize the discriminating value of a term over its descriptive value. Alternatively, if the user is seeking to satisfy a more general information needs by looking for as many relevant results as possible, then we expect recall to be favored by selecting terms with high \fddscore for high $\beta$ values. In this case, good descriptors should be prioritized over good discriminators. The analysis and evaluations presented in the next section provide empirical evidence showing that this intuition is indeed correct.
    
    \vspace{-0.5cm}

\section{The role of  $\beta$ in \fddscore }
\label{sec:analysis}

    The behavior of  \fddscore as a function of $\beta$
    and its effectiveness as a query term selection technique was analyzed on three data sets, which are described next.  The code used to carry out this analysis is made available to the research community for reproducibility.\footnote{\url{http://cs.uns.edu.ar/~mmaisonnave/resources/FDD_code}.}
    
    \subsection{Data sets} \label{sec:datacollection}
    The first data set used to analyze the role of $\beta$, to which we refer to as {\em ERNTG}  ({\em Economic Relevant News from The Guardian}), was labeled by experts in Economy and consists of news articles collected from the Politics, World news, Business and Society sections of the  {\em The Guardian} newspaper (\url{https://www.theguardian.com/}). A set of 1689 news articles corresponding to January 2013 were collected using a Python script that calls an API provided by the newspaper. Although several fields are available for each news article, only the news titles and full body text were used. Texts were processed using the Spacy NLP library version 2.1.4 (\url{https://pypi.org/project/spacy/}) with the default setting. Each news article was tokenized and each token was lemmatized. Hyphenated terms are considered as independent terms. Finally, the vocabulary was built using  the identified tokens, excluding  irrelevant terms such as stopwords, terms that are infrequent in the data set (occurring in less than 15 articles), and words with non-alphabetic characters. 
     To create the training set  two experts in Economy read the collected 1689 news articles and agreed on whether each of them was relevant or not to the economic domain. As a result, 537 of the articles were marked as relevant and 1152 as irrelevant.  It is worth mentioning that the manual labeling stage was important due to the fact that  news identified by the experts as  economic relevant do not exactly correspond to those from the Business section (418 out of 512) but also some of them were in the Politics (39 out of 290), World news (43 out of 650), and Society (37 out of 237)  sections.  Another 100 expert-labeled news articles were used for testing. We have made the data set available for reproducibility  and its usage for testing other methods~\citep{maisonnave2019news}.\footnote{\href{http://dx.doi.org/10.17632/yt8j2f3hpp.3}{http://dx.doi.org/10.17632/yt8j2f3hpp.3}.}

    The second data set is the  widely used  {\em 20 Newsgroups}, also known as {\em 20NG}  (\url{http://qwone.com/~jason/20Newsgroups/}), which consists of approximately 20000 newsgroup documents, from 20 different newsgroups, each corresponding to a different category. Some of the newsgroups are very closely related to each other (e.g. comp.sys.ibm.pc.hardware and comp.sys.mac.hardware), while others are highly unrelated (e.g. misc.forsale and soc.religion.christian). In the same way as for the {\em ERNTG} data set, the vocabulary was built by processing the documents using the  Spacy NLP library. The data set was split into 80\% for training and 20\% for testing. 
    
    The third data set used is the {\em Reuters-21578} Text Categorization Test Collection, also known as {\em Reuters} (\url{https://archive.ics.uci.edu/ml/data sets/reuters-21578+text+categorization+collection}). The data set consists of 21578 data items collected from the Reuters newswire in 1987. The documents were assembled and indexed with categories by personnel from Reuters Ltd.  Each document could have zero or more associated categories (or topics), from a total of 120 topics. Not all the data items in the data set have a full text associated with them. Therefore, the final data set consists of 19043 text documents, each document having zero or more of the 120 possible topics. Some examples of the topics are {\em housing}, {\em livestock} and {\em jobs}. In the same way as for the other data sets, we performed the pre-processing of the data set using the Spacy NLP library. The data set was split into 80\% for training and 20\% for testing. We do not use this third data set for analyzing the role of $\beta$ in \fddscore. Instead, we use it as an additional data set for the evaluation of retrieval effectiveness.

    \vspace{-0.2cm}

     \subsection{Analysis on the ERNTG data set}
     
     In order to analyze the behavior of the parameter $\beta$  we looked into the  retrieval performance of queries generated using terms selected based on the best \fddscore for different values of $\beta$.
     The evaluated queries were generated using the unigrams, bigrams and trigrams with the highest \fddscore scores on the training portion of the  {\em ERNTG} data set based on $\beta$ values ranging from 0 to 10.
     Using the resulting queries, we computed the classical \precision, \recall and \fonescore metrics on the training and test sets.
      Table~\ref{table:besttermsERNTG} illustrates the role of the $\beta$ parameter by showing the $\beta$ ranges and performance values. Each $\beta$ range represents the interval for which the set of terms with the highest \fddscore on the training set remains invariant. For instance, the bigram ``Royal Bank''  achieves the highest  \fddscore on the training set for $\beta \in [0.07,0.14)$. This query achieves  \recall values of 0.09 (training set) and 0.07 (test set),   \precision value of 0.97 (training set) and 1.00 (test set) and an \fonescore of 0.16 (training set) and 0.13 (test set).

     \begin{table}
     	\centering
     	{\footnotesize
     	\renewcommand{\arraystretch}{0.7}
     	\begin{tabular}{llccc}
     		\hline
     		Term &        $\beta$ range           &   \recall   & \precision   & \fonescore\\
     		\hline
                Capital Economics & [0.00,0.01) & 0.04/0.03 & 1.00/1.00 & 0.08/0.06 \\
                bank say & [0.01,0.04) & 0.05/0.04 & 1.00/1.00 & 0.10/0.07 \\
                Federal Reserve & [0.04,0.07) & 0.05/0.07 & 1.00/1.00 & 0.10/0.13 \\
                Royal Bank & [0.07,0.14) & 0.09/0.07 & 0.97/1.00 & 0.16/0.13 \\
                economist & [0.14,0.24) & 0.17/0.19 & 0.88/0.83 & 0.29/0.31 \\
                investor & [0.24,0.36) & 0.26/0.25 & 0.81/0.74 & 0.39/0.38 \\
                growth & [0.36,0.49) & 0.39/0.41 & 0.71/0.70 & 0.51/0.52 \\
                market & [0.49,1.25) & 0.50/0.43 & 0.65/0.59 & 0.57/0.50 \\
                year & [1.25,1.34) & 0.84/0.89 & 0.36/0.36 & 0.51/0.51 \\
                have & [1.34,10.0]& 0.99/1.00 & 0.32/0.31 & 0.49/0.47 \\
                \hline
    \end{tabular}\\	}
     		\caption{Terms maximizing \fddscore learned from the training set for different $\beta$ ranges and their performance as queries on the training/test sets   ({\em ERNTG} data set).}
     		 	\label{table:besttermsERNTG}
     	\end{table}

\begin{figure}[H]
     	 		\vspace*{-0.4cm}	
     	 	\hspace*{-1.3cm}
     	 	\includegraphics[width=1.2\textwidth]{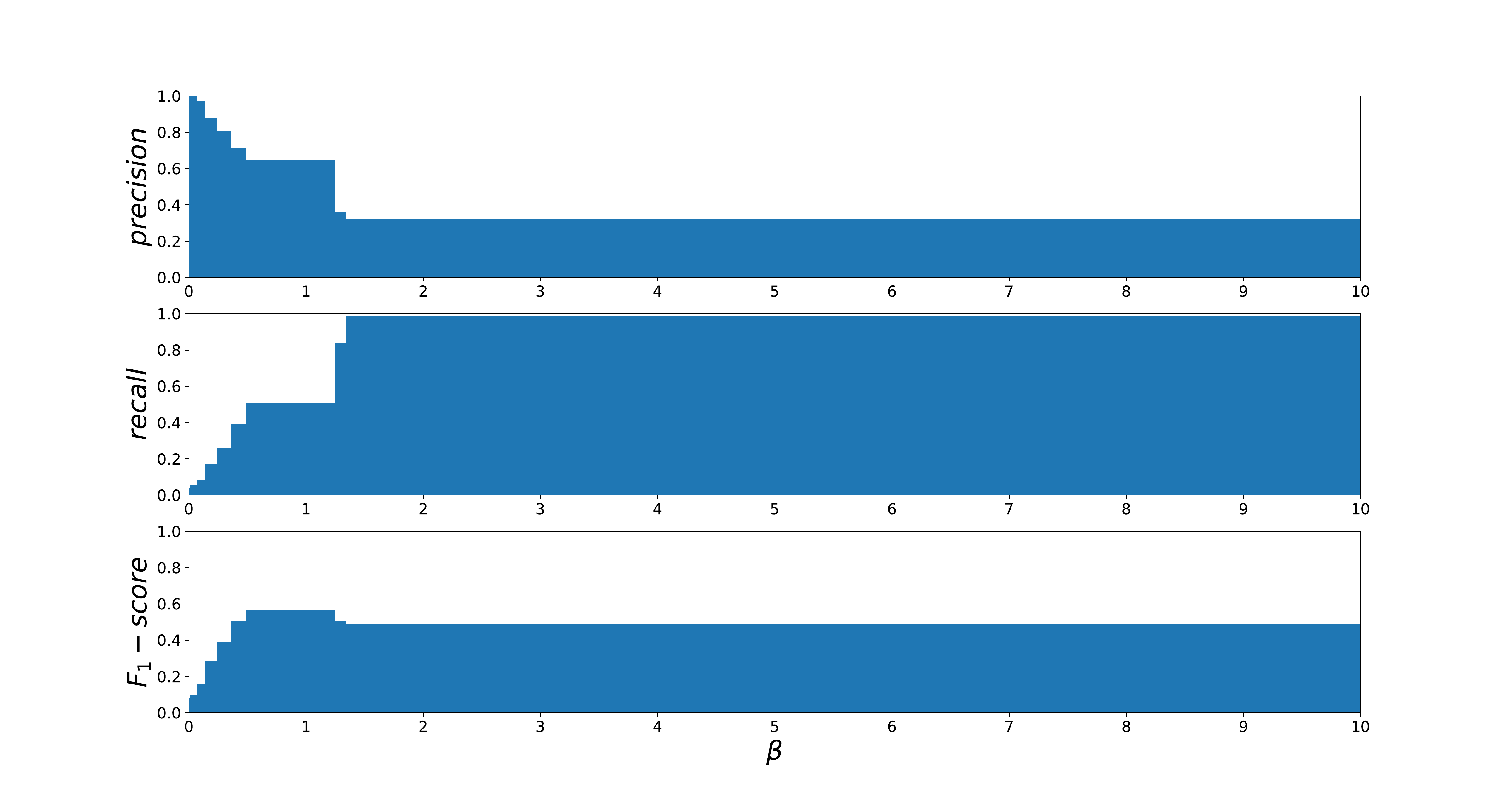}
     	 		 	\vspace*{-1.3cm}
     	 	\caption{ Effectiveness on the training  set of best query terms selected from the training set using \fddscore  with different  $\beta$ values ({\em ERNTG} data set).}
     	 	\label{fig:theroleofbetaERNTGTraining}
     	 \end{figure}

\begin{figure}[!ht]	
     	   		\vspace*{-0.4cm}
     	   	\hspace*{-1.3cm}
     	   	\includegraphics[width=1.2\textwidth]{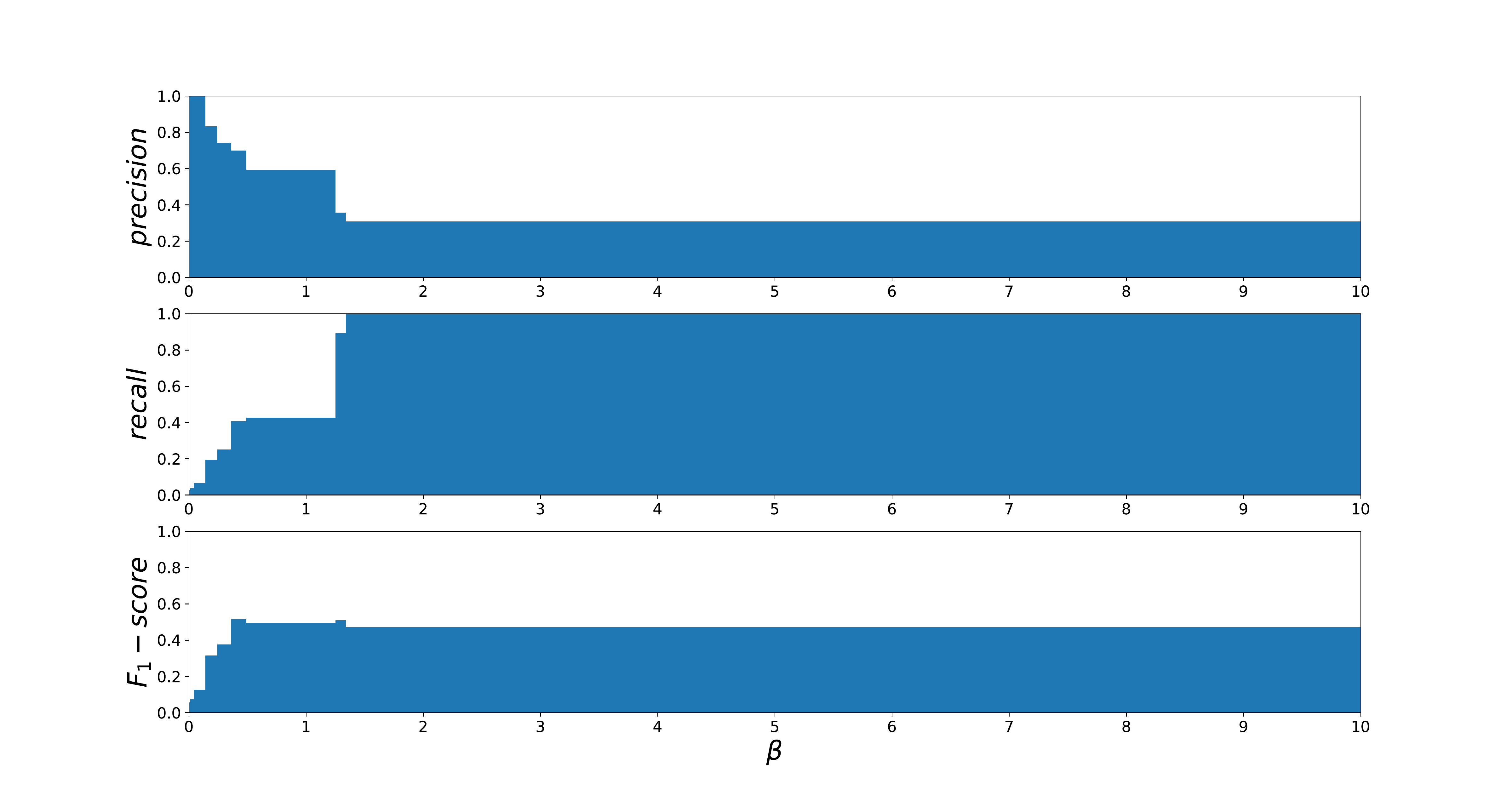}
     	   		 	\vspace*{-1.3cm}
     	   	\caption{ Effectiveness on the test set of best query terms selected from the training set using \fddscore  with different  $\beta$ values ({\em ERNTG} data set).}
     	   	
     	   	\label{fig:theroleofbetaERNTGTesting}
     	   \end{figure}

     	To visualize the dynamics of \precision, \recall and \fonescore for different $\beta$ values, figure~\ref{fig:theroleofbetaERNTGTraining}  shows the performance  achieved by those   queries (based on unigrams, bigrams and trigrams) maximizing \fddscore  learned from the training set for different values of $\beta$ ranging from 0 to 10. 
     	The analysis of these results shows that the $\beta$ parameter has a crucial effect on retrieval performance. As anticipated, we can observe that low  $\beta$ values favor \precision and high $\beta$ values favor \recall.  The optimal $\beta$ for a specific performance metric value can be learned from the training set. Our analysis  presented in figure~\ref{fig:theroleofbetaERNTGTesting} shows  the performance on the test set  of the   queries maximizing \fddscore identified using the training set. It is possible to verify that the behaviors observed on the training and test sets are similar, indicating that the analyzed term selection strategies based on \fddscore do not suffer from overfitting for the analyzed setting.

     	 \vspace{-0.2cm}
     	 
     	 \subsection{Analysis on the 20NG data set}
     	 
\begin{figure}[!htb]
    		
    	\hspace*{-1.3cm}
    	\includegraphics[width=1.2\textwidth]{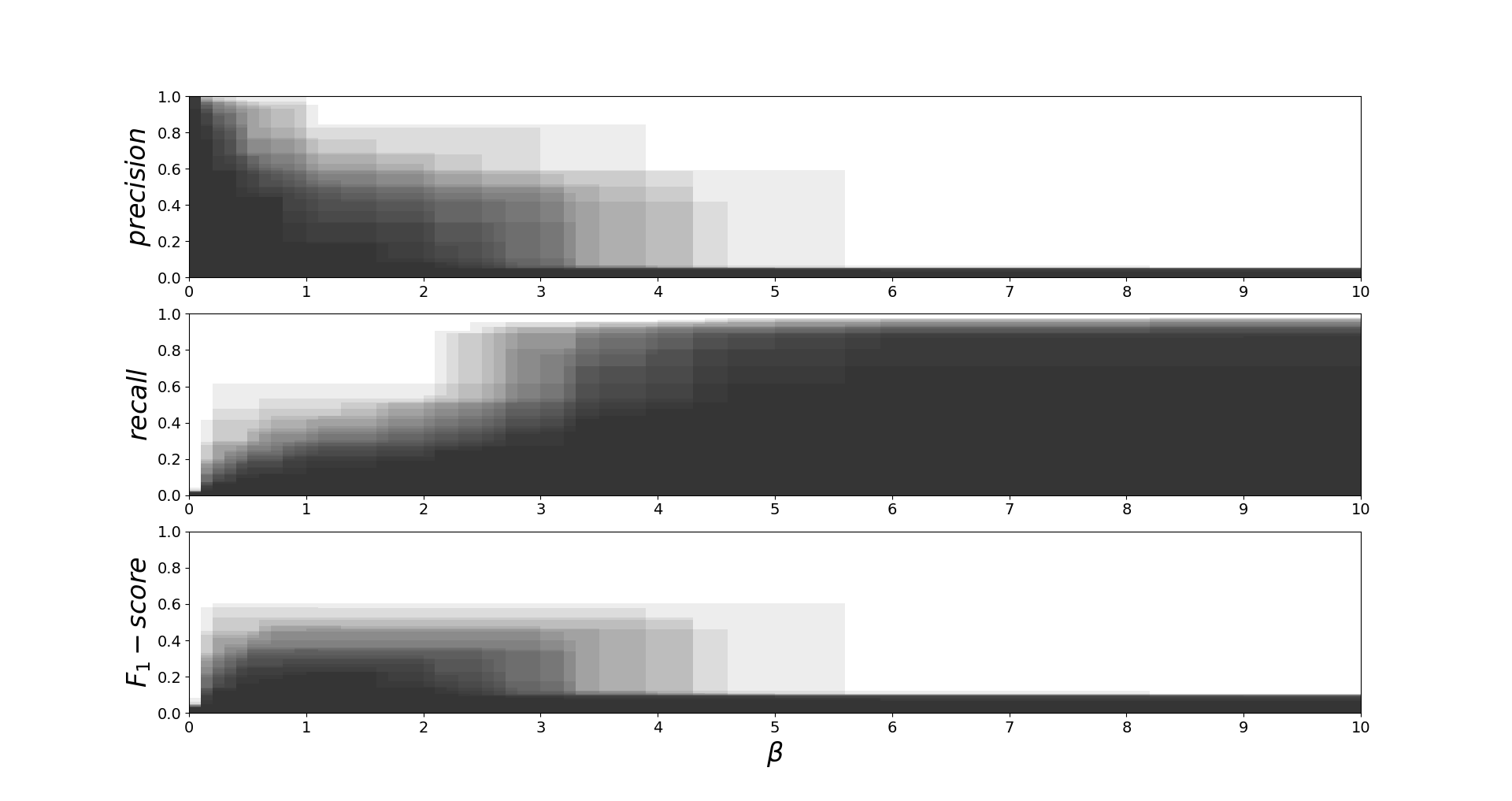}
    	\vspace*{-1.3cm}
    	\caption{ Effectiveness on the training  set of best query terms selected from the training set using \fddscore  with different  $\beta$ values. Each shadow represents a different category ({\em 20NG} data set).}
    	\label{fig:theroleofbeta20NGTraining}
    \end{figure}

\begin{figure}[!htb]	
    	\hspace*{-1.3cm}
    	\includegraphics[width=1.2\textwidth]{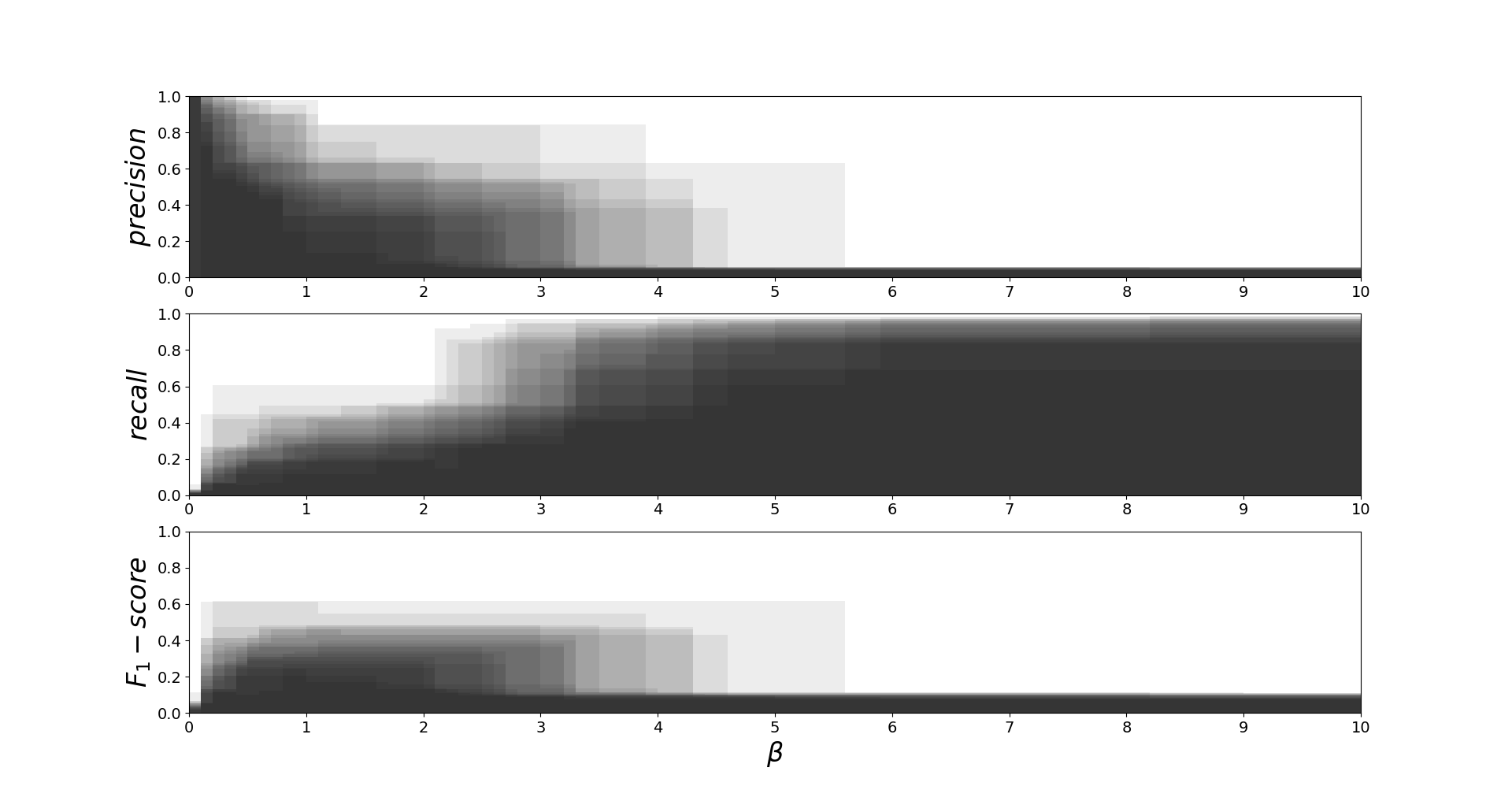}
    	\vspace*{-1.3cm}
    	\caption{ Effectiveness on the test   set of best query terms selected from the training set using \fddscore  with different  $\beta$ values.  Each shadow represents a different category ({\em 20NG} data set).}
    	\label{fig:theroleofbeta20NGTesting}
    \end{figure}

\begin{figure}[!ht]	
    	\hspace*{-1.3cm}
    	\includegraphics[width=1.2\textwidth]{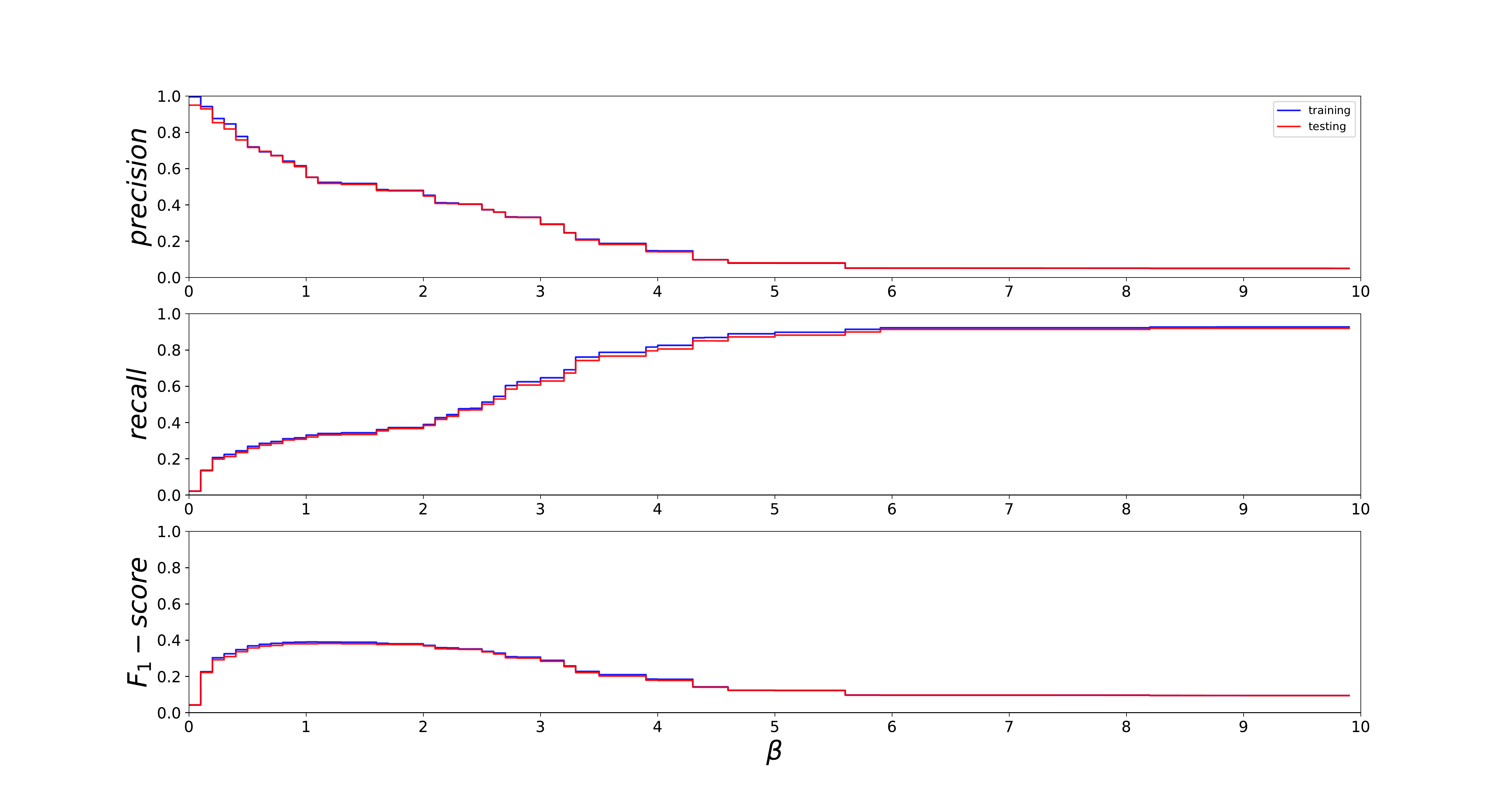}
    		\vspace*{-1.3cm}
    	\caption{ Effectiveness on the training  and test sets of best query terms selected from the training set using \fddscore  with different  $\beta$ values averaged across the 20 categories ({\em 20NG} data set).}
    	\label{fig:theroleofbeta20NGAveragedTrainingAndTesting}
    \end{figure}

\begin{figure}[!ht]	
    	\hspace*{0cm}
    	\includegraphics[width=1.0\textwidth]{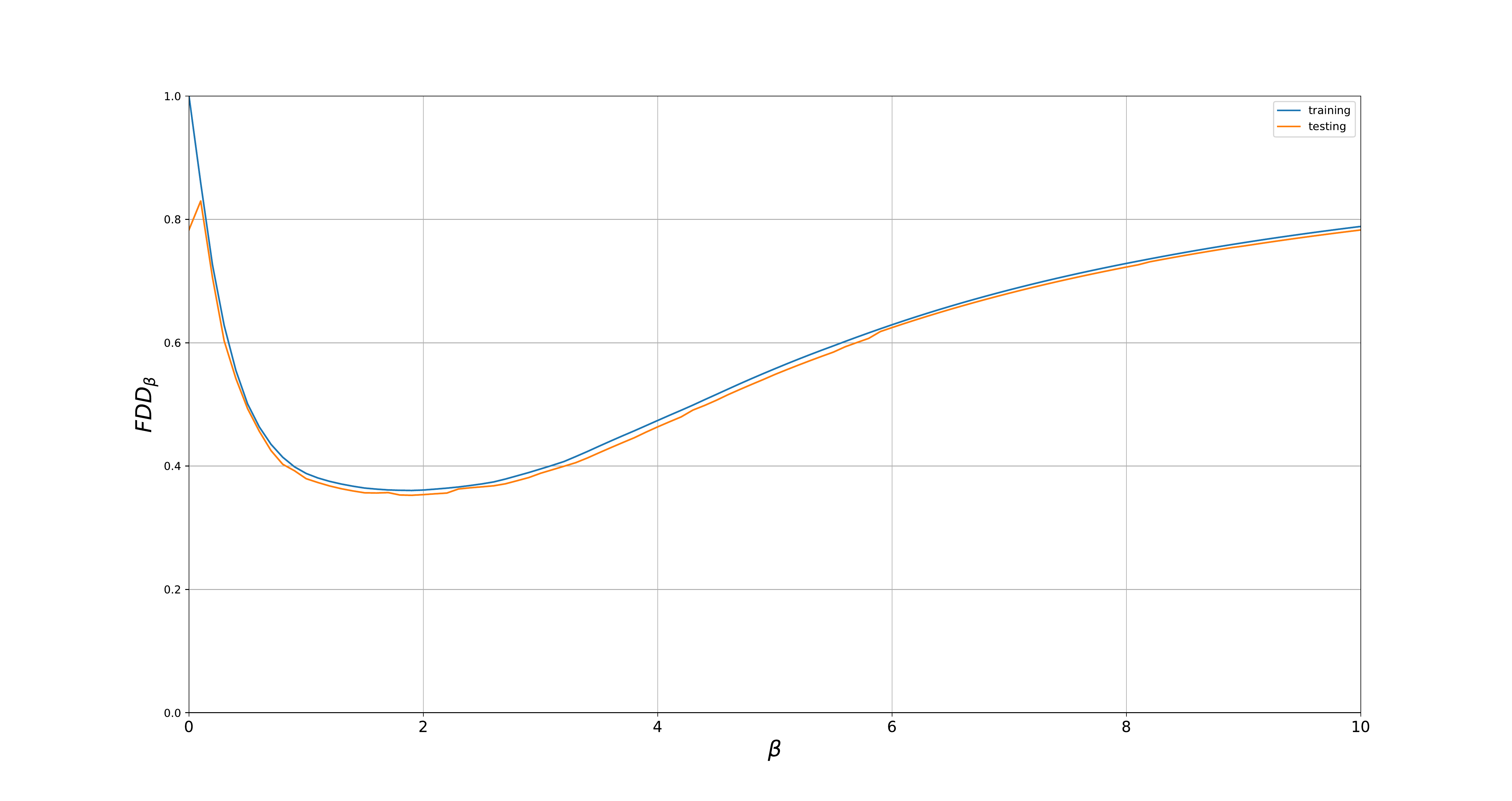}
    		\vspace*{-1.3cm}
    	\caption{ \fddscore on the training and test sets of best query terms selected from the training set using \fddscore  with different  $\beta$ values ({\em 20NG} data set).}
    	\label{fig:fddbetaforbeta}
    \end{figure}

     	 A similar but more extensive analysis was completed for the {\em 20NG} data set. For each of the 20 categories, the terms (unigrams, bigrams or trigrams) maximizing  \fddscore  were identified at different values of $\beta$. 
     	 The performance of the selected terms when used as query terms was evaluated  on the training and test sets. 
     	  A visualization of the dynamics of \precision, \recall and \fonescore for different $\beta$ values on the training and test sets are presented in figures~\ref{fig:theroleofbeta20NGTraining} and~\ref{fig:theroleofbeta20NGTesting}, respectively. In these figures, each shadow represents a category. 
      The averaged performance values across the 20 categories are shown in figure~\ref{fig:theroleofbeta20NGAveragedTrainingAndTesting}. 
      Once again, it is possible to verify that  low $\beta$ values favor precision while high $\beta$ values favor recall.   Finally, figure~\ref{fig:fddbetaforbeta} shows the \fddscore values on the training and test sets of the best query terms selected from the training set using \fddscore  with different  $\beta$ values. The similar performance achieved on the training and test sets indicates, once again, that the term selection strategies based on \fddscore do not suffer from overfitting for the analyzed setting.

    \newpage
    
    To dig deeper into the analysis of  the $\beta$ parameter, we looked into the question of which are the terms (unigrams, bigrams or trigrams) that result in the highest \fonescore on the training set for each of the 20 categories in the {\em 20NG} data set. Table~\ref{table:bestBeta} reports these results, presenting for each category the $\beta$ ranges maximizing  \fonescore, a representative query term and the \fonescore value achieved on the training and test sets.  Tables with representative  query terms selected based on the best \fddscore for different ranges of $\beta$ and their performance in terms of \precision, \recall and \fonescore  on the training and test sets are available as part of the supplementary material.\footnote{\href{https://cs.uns.edu.ar/~mmaisonnave/resources/maisonnave2020FDD-supplementary.pdf}{https://cs.uns.edu.ar/$\sim$mmaisonnave/resources/maisonnave2020FDD-supplementary.pdf}}
    
      \begin{table}
      	\centering
      	{\footnotesize
      	\renewcommand{\arraystretch}{0.7}
      	\begin{tabular}{llll }
      		\hline
      		 	Category &        $\beta$ range           &   query term   &  \fonescore (training/testing) \\
      		\hline
      		
      alt.atheism       &        [0.3,1.9)   &  atheist & 0.32  / 0.29 \\
      comp.graphics             & [0.8,2.0) & image  & 0.27 / 0.25  \\
      comp.os.ms-windows.misc   & [0.2,4.2) &  Windows & 0.53 / 0.47 \\
      comp.sys.ibm.pc.hardware  & [0.8,2.5)   & card & 0.30 / 0.33 \\
      comp.sys.mac.hardware     & [0.3,2.4) & Apple & 0.36 / 0.37 \\
      comp.windows.x            & [0.5,3.2) & X & 0.40 / 0.43 \\
      misc.forsale              & [0.4,3.1)  & sale & 0.35 / 0.37 \\
      rec.autos                 & [0.2,5.5)  & car & 0.60 / 0.62 \\
      rec.motorcycles           & [0.1,1.0) & bike  & 0.58 / 0.61 \\
      rec.sport.baseball        & [0.5,1.0) & pitch  & 0.35 / 0.31 \\
      rec.sport.hockey          & [0.7,1.2)  & team & 0.48 / 0.46 \\
      sci.crypt                 & [1.0,3.4)  & key & 0.46 / 0.48 \\
      sci.electronics           & [0.1,1.5)  & circuit & 0.25 / 0.20\\
      sci.med                   & [0.4,2.0)  & doctor  & 0.30 / 0.31 \\
      sci.space                 & [0.9,2.6)  & space &  0.36 / 0.34 \\
      soc.religion.christian    & [0.6,4.2) &  God & 0.51 / 0.46 \\
      talk.politics.guns        & [0.5,3.1)  & gun & 0.45 / 0.38 \\
      talk.politics.mideast     & [0.6,2.9)  & Israel & 0.48 / 0.48\\
      talk.politics.misc        & [1.0,1.6)  & government & 0.23 / 0.17 \\
      talk.religion.misc        & [0.8,2.6)&  Christian & 0.23 / 0.27 \\ \hline

     \end{tabular}
     }
     \caption{Best $\beta$ value intervals for the \fonescore measure  learned from the training set of each  category, selected query terms and resulting \fonescore on training/test sets ({\em ERNTG} data set). }
     \label{table:bestBeta}
    \end{table}

\section{Evaluation of retrieval effectiveness}
\label{sec:evaluation}
    
    In this section, we analyze the performance of \fddscore as a mechanism for query-term selection in topic-based retrieval and we compare it with the eighteen traditional and state-of-the-art weighting schemes described in table~\ref{table:definition}.  The code used to carry out the reported evaluations is made available through Google Colab to the research community to facilitate reproducibility.\footnote{\url{http://cs.uns.edu.ar/~mmaisonnave/resources/FDD_code}.} 
    
    The comparative evaluation was completed on the  {\em 20NG} and {\em Reuters} data sets. For both {\em 20NG} and {\em Reuters}, we use the training portion for selecting the top-rated terms (unigrams, bigrams or trigrams) for each category, using all the schemes described in table~\ref{table:definition}. To illustrate how  \fddscore can favor different retrieval goals, we report results for  \fddscore with (1) $\beta = 0.5$ to favor precision over recall, (2) $\beta = 10$ to favor recall over precision to a large degree, and  (3) $\beta = 1$ to  equally balance precision and recall.
    Simple queries consisting of a single  unigram, bigram or trigram were generated using the selected terms according to each weighting  schemes and then evaluated by means of the classical  \precision, \recall and \fonescore metrics on the training and test sets.  
     
     
     \subsection{Evaluation on the 20NG data set}
\begin{figure}[!ht]	
    	\hspace*{-1.3cm}
    	\includegraphics[width=1.2\textwidth]{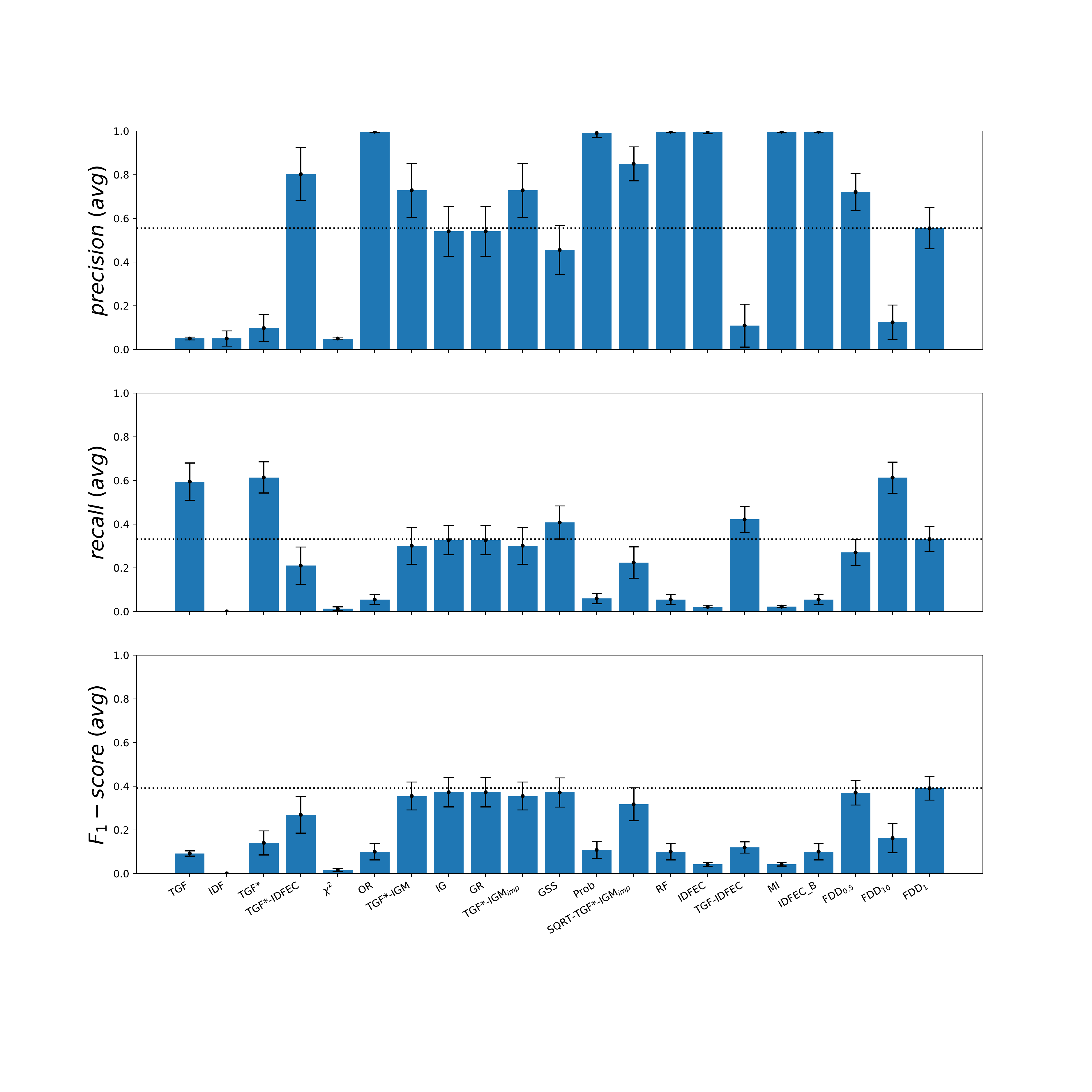}
    	\vspace*{-1.5cm}
    	\caption{ Performance comparison of the analyzed methods on the training set  ({\em 20NG} data set - averaged across the 20 categories).}
    	\label{fig:performanceComparison20NGTraining}
    \end{figure}

\begin{figure}[!ht]	
    	\hspace*{-1.3cm}
    	\includegraphics[width=1.2\textwidth]{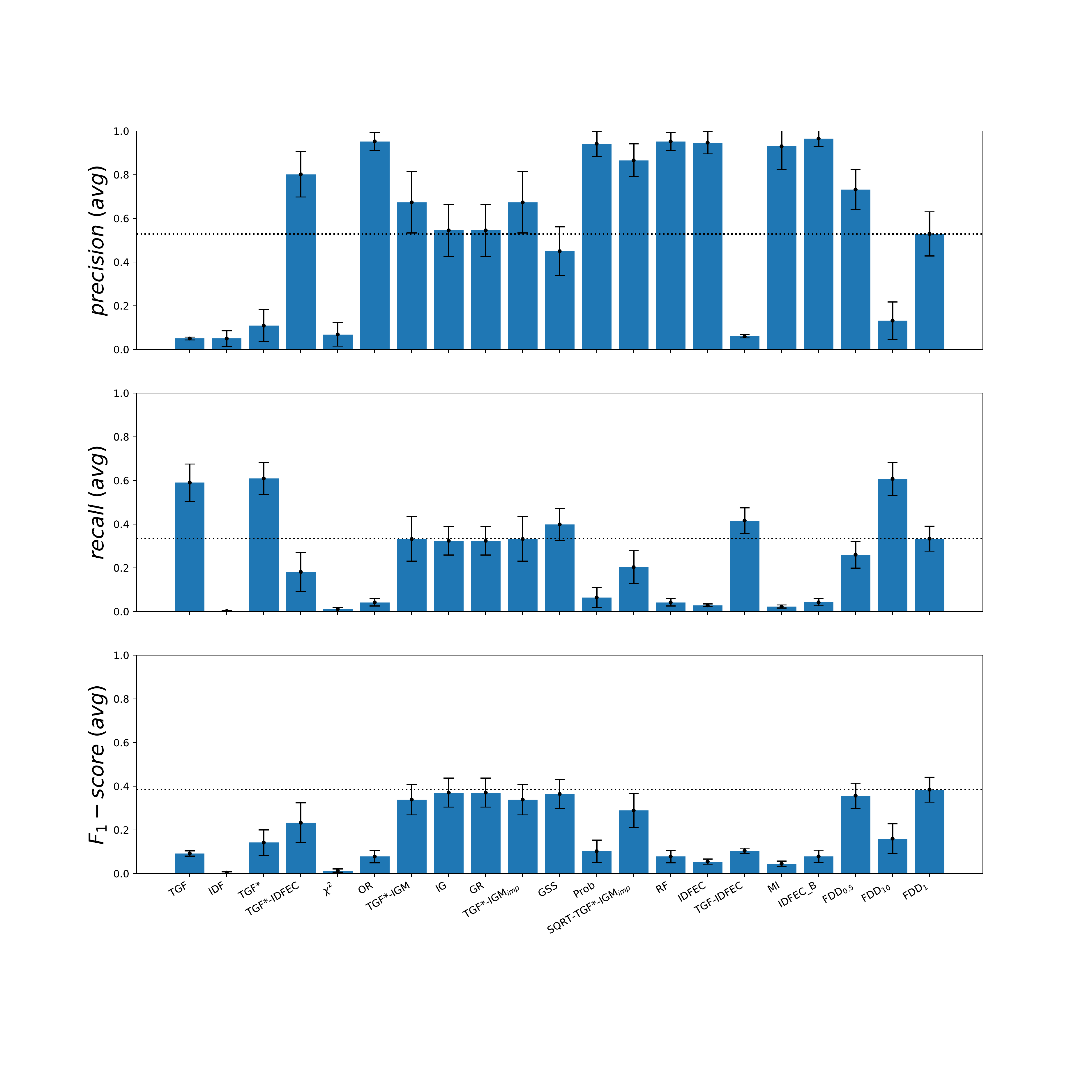}
    	\vspace*{-1.5cm}	
    	\caption{ Performance comparison of the analyzed methods on the test set   ({\em 20NG} data set - averaged across the 20 categories).}
    	\label{fig:performanceComparison20NGTesting}
    \end{figure}

     In this section, we present the results and the discussion of the experiments performed on the {\em 20NG} data set. 
     The results of the comparative analysis for the training and test sets are shown in figures~\ref{fig:performanceComparison20NGTraining} and ~\ref{fig:performanceComparison20NGTesting}, respectively. Based on this evaluation, the \fddscore scheme with $\beta = 1$ achieves an \fonescore score  higher than that obtained by all the other techniques.
     In particular, it is higher than the one obtained with the most effective weighting schemes, namely TGF*-IGM, IG, GR, TGF*-IGM$_{imp}$, GSS and SQRT-TGF*-IGM$_{imp}$, both on the training and test sets (although the improvements are not statistically significant). Moreover,   IG, GR and GSS are outperformed in terms of precision  by   \fddscore with $\beta = 0.5$ and in terms of recall by  \fddscore with $\beta = 10$.  Also, while some methods, such as OR, Prob, RF, IDFEC, MI, and IDFEC\_B are very effective when evaluated in terms of precision, they perform poorly in terms of recall and \fonescore. Conversely, other methods such as TGF and TGF* achieve the highest recall, but have a poor performance in terms of precision and \fonescore.  Finally, it is worth mentioning that unsupervised methods that put high emphasis on favoring highly discriminative terms, such as IDF and $\chi^2$ perform badly both on the training and test sets. This is due to the fact that these methods tend to select rare terms from the training set independently of the relevant class. These results point to the importance of taking advantage of labeled data in  the process of learning good query terms as well as to the usefulness of adjusting the term-weighting scheme to the task goals, as can be naturally done by means of the  $\beta$ parameter in the \fddscore.  
     It is worth noticing that all the IGM-based methods have a good performance. These results suggest that combining the class distinguishing power of a term  with a term frequency factor is a promising direction for a term-weighting technique. This conclusion further supports the intuition behind the  \fddscore term-weighting technique.

\begin{figure}[!ht]
     	\hspace*{-1.3cm}
     	\includegraphics[width=1.2\textwidth]{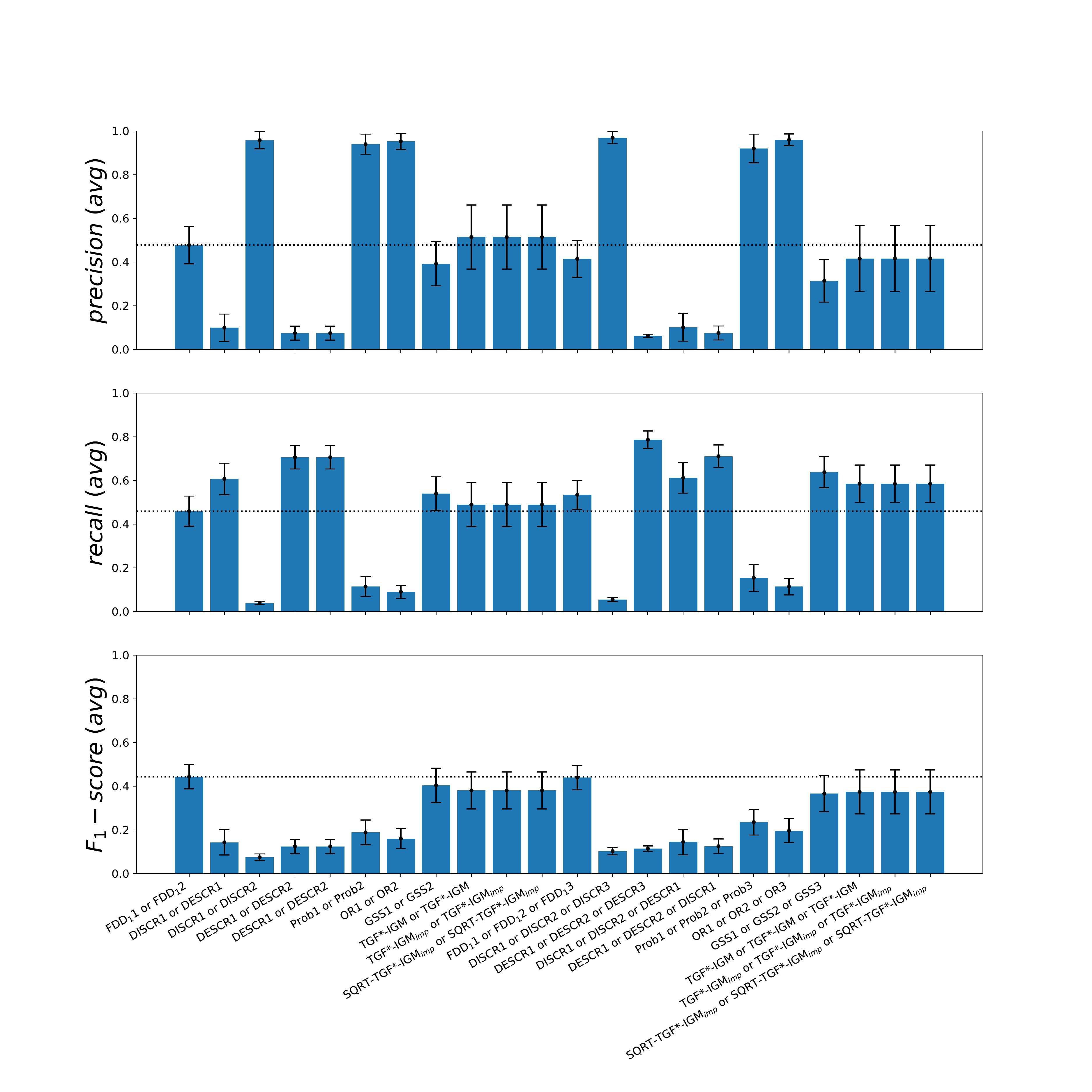}
     	\hspace*{-1.3cm}
     		\vspace*{-1.5cm}	
     	\caption{ Performance comparison on the test set of complex queries generated with terms selected using different methods  ({\em 20NG} data set - averaged across the 20 categories).}
     	\label{fig:complexqueries}
     \end{figure}

    The previous analysis relies  on the use of simple queries (i.e., queries consisting of single unigrams, bigrams or trigrams). More complex queries can be formed by introducing Boolean operators such as ``and'' and ``or''.   As mentioned earlier, a difficulty in dealing with query optimization is  that optimal complex queries cannot be constructed efficiently from optimal simple queries. As a consequence, combining good query terms may not necessarily result in an effective longer query. While the analysis of complex queries could involve many aspects and dimension, for the sake of simplicity, we limit the analysis to disjunctive queries of size two and three. This choice is also guided by previous studies that indicate that disjunctive  queries tend to be more effective than conjunctive queries~\citep{cecchini09evolving}. Figure~\ref{fig:complexqueries} presents the results achieved by disjunctive queries built by combining terms selected using some of the analyzed methods. For instance, query ``FDD$_1$1 or  FDD$_1$2'' stands for the disjunctive  query built using the two top-ranked terms (unigrams, bigrams or trigrams) based on the  FDD$_1$ weighting scheme. Similarly, ``DISCR1 or DESCR1'' represents the query built by a disjunction of the best discriminator (DISCR1) and the best descriptor (DESCR1), and so on. 
    
    As expected, the reported results highlight the advantage of using highly discriminative terms (e.g. ``DISCR1 or DISCR2'' and ``DISCR1 or DISCR2 or DISCR3'') to achieve high \precision, while good descriptors  (e.g. ``DESCR1 or DESCR2'' and ``DESCR1 or DESCR2 or DESCR3'') allow to achieve high \recall. Finally, we can observe that the best \fonescore is achieved by selecting terms with high FDD$_1$ score (e.g. ``FDD$_1$1 or FDD$_1$2'' and ``FDD$_1$1 or FDD$_1$2 or FDD$_1$3'').
    
    Notice that for the FDD$_1$ scheme, the two-term queries (FDD$_1$1 or FDD$_1$2) have slightly better performance than the three-term queries (FDD$_1$1 or FDD$_1$2 or FDD$_1$3). This points to the fact that  incorporating query terms selected based on the  FDD$_\beta$ scheme with the same $\beta$ value is not always beneficial. This result shows that more complex queries may be required, and the analyzed technique with a tunable parameter is a good candidate to explore different approaches to build complex queries.

     \subsection{Evaluation on the Reuters data set}
    
\begin{figure}[!ht]	
            	\hspace*{-1.3cm}
            	\includegraphics[width=1.2\textwidth]{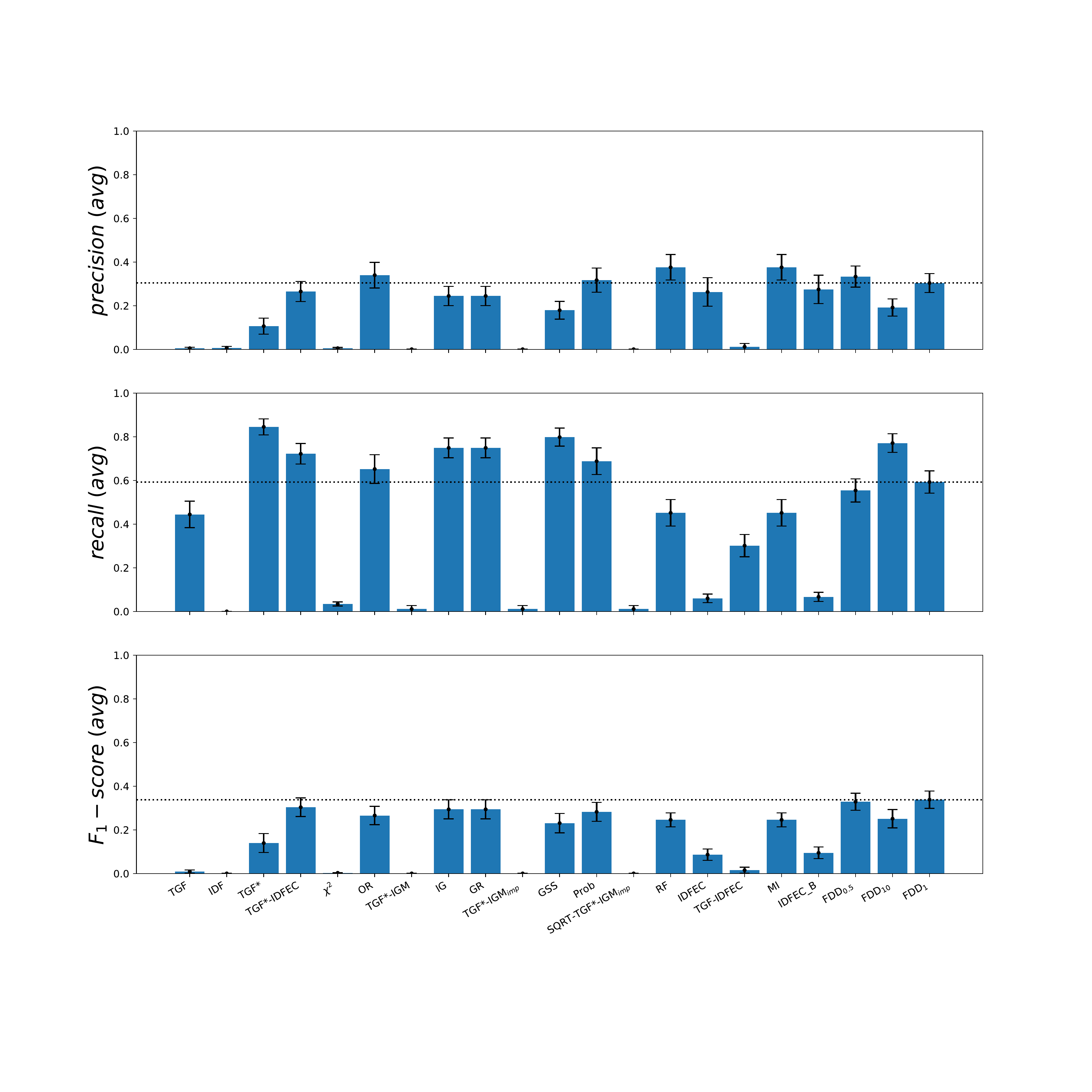}
            	\vspace*{-1.5cm}
            	\caption{ Performance comparison of the analyzed methods on the training set  ({\em Reuters-21578} data set - averaged across the 120 topics).}
            	\label{fig:performanceComparison20ReutersTraining}
            \end{figure}

\begin{figure}[!ht]	
            	\hspace*{-1.3cm}
            	\includegraphics[width=1.2\textwidth]{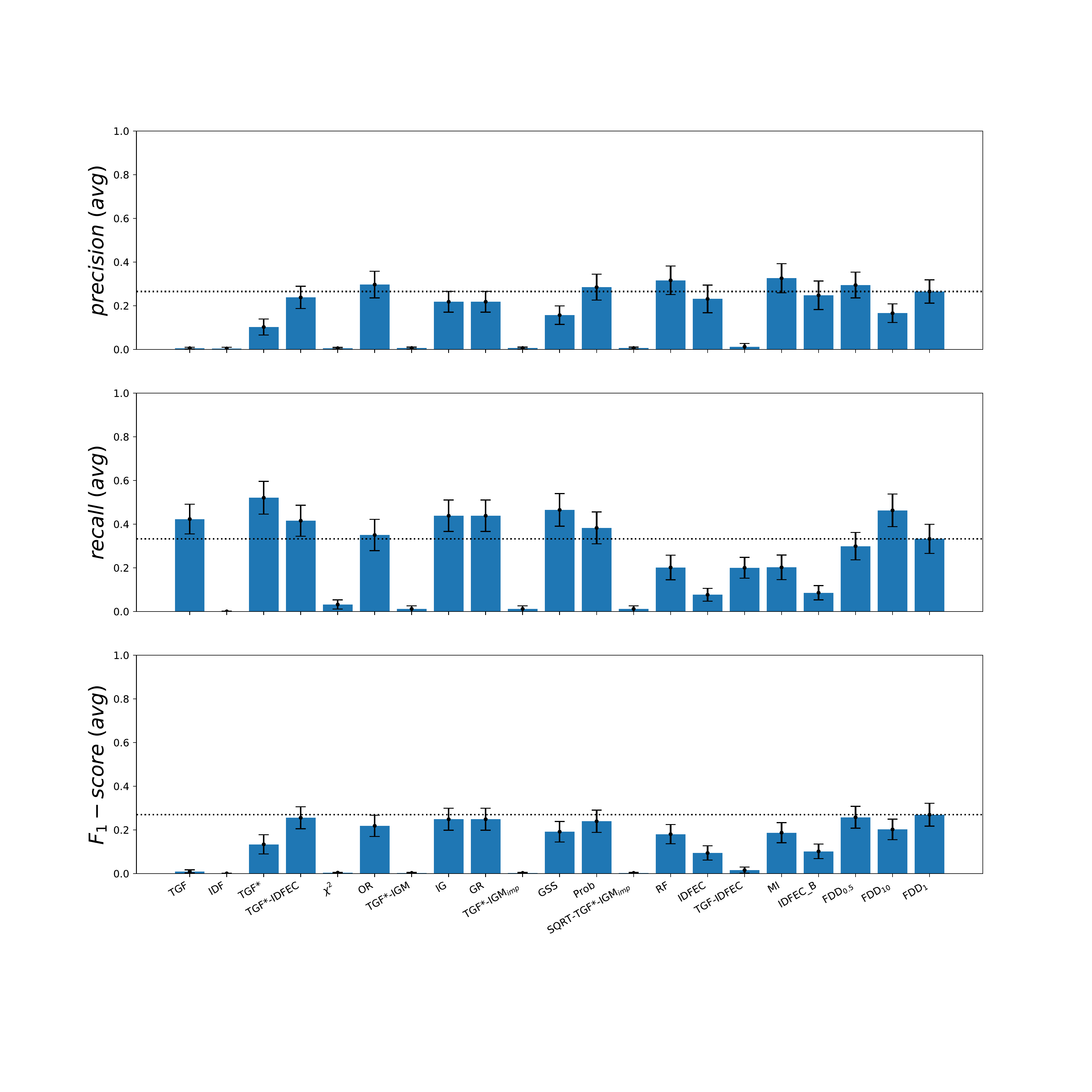}
            	\vspace*{-1.5cm}	
            	\caption{ Performance comparison of the analyzed methods on the test set   ({\em Reuters-21578} data set - averaged across the 120 topics).}
            	\label{fig:performanceComparisonReutersTesting}
            \end{figure}

     In this section, we present and discuss the results obtained from comparing the \fddscore scheme with other state-of-the-art schemes using the {\em Reuters} data set. To carry out the evaluations we used the training set to select the top-ranked terms based on each term-weighting method. The selected terms were used to build single-term queries in the same way as we did with the {\em 20NG} data set. The performance of each evaluated method based on the classic metrics of \precision, \recall and \fonescore on both the training and test sets are reported in figures \ref{fig:performanceComparison20ReutersTraining} and \ref{fig:performanceComparisonReutersTesting}, respectively.

    Note that in the {\em Reuters} data set, the scenario is slightly different due to the characteristics of the categories. In this data set there are 120 categories, and each document has zero or more associated topics. While in the {\em 20NG} data set, each document has only one associated category. Because of this setting in {\em 20NG}, we were able to build a richer class distribution (of dimension 20), for computing all the IGM-based methods. In the case of the {\em Reuters} data set,  we had to simplify the problem to binary categories (belonging to the given topic or not) for computing the class distribution due to the non-exclusivity of the labels.

    Once again, the results show that \fddscore achieves the best \fonescore (although the improvements are not always statistically significant).  It is worth mentioning that some methods that  achieved a good performance in the {\em 20NG} dataset, such as IG, GR and GSS, are consistently effective in the {\em Reuters} data set. On the other hand, the methods based on IGM, which were among the most effective ones using the {\em 20NG} data set, perform poorly in the {\em Reuters} data set. The poor performance of the IGM-based techniques is a consequence of the aforementioned  simplification from a class distribution of 20 values to one of only 2. This indicates that the IGM-based methods are not suited for this binary setting.

     We also observe that FDD$_{0.5}$ is slightly inferior to MI and RF in terms of \precision but superior to both of these methods in terms of \recall and \fonescore.  Finally, we observe that FDD$_{10}$ achieves a \recall value that is comparable, although slightly inferior, to the \recall achieved by TFG*, which is the state-of-the-art method that achieved the highest \recall value. However, all the analyzed \fddscore schemes are superior to  TFG* in terms of \precision and \fonescore. Note that, in general, the performance  achieved for \fonescore, \precision and \recall in the {\em Reuters} data set was systematically lower for all the evaluated methods compared to the experiments on the {\em 20NG} data set.

\section{Discussion}
\label{sec:discussion}

The analysis presented in section \ref{sec:analysis} allowed to take a close look into the behavior of \fddscore as a function of the parameter $\beta$. We observed that the $\beta$ parameter has a crucial effect on retrieval performance and that its optimal value can be learned from the training set based on the retrieval goal (high precision, high recall or a balance of both).  We observed that the performance of \fddscore was similar on the training and test sets, indicating that it does not suffer from overfitting for the analyzed settings. 

The extensive evaluations reported in section~\ref{sec:evaluation} show that the \fddscore scheme with $\beta$=1 consistently achieves an \fonescore score higher than the one obtained by all the other evaluated techniques, both on the training and testing sets. Also, the parameter $\beta$ can be adjusted to outperform most of the evaluated techniques in terms of \precision and \recall. The evaluations also shed some light on the advantages of supervised approaches to term-weighting schemes when compared to unsupervised ones. It is  important to highlight that \fddscore with $\beta$=1 also resulted in the best \fonescore score when multi-term disjunctive queries were evaluated. Finally, we demonstrated the usefulness of generating multi-term queries based on descriptors or discriminators to attain high recall or precision, respectively.

A major finding of our research is that despite the fact that information-theoretic and statistical notions such as entropy, mutual information, information gain and probability distributions have shown to be especially useful in the definition of unsupervised term-weighting methods, we contend that the level of analysis adopted by these schemes is not necessarily required for supervised term-weighting approaches in the topic-based retrieval task. As can be seen from our results, \fddscore is sufficient to achieve highly competitive results without the need to rely on the aforementioned notions. The results derived from this research demonstrate that the \fddscore weighting scheme offers a useful mechanism to explore different approaches to build complex queries for topic-based retrieval. 

\section{Conclusions and future work }
\label{sec:conclusion}

    
    
    This article analyzed the behavior of the \fddscore term-weighting score and evaluated it performance in query-term selection for topic-based retrieval. 
    The first contribution of this work is an extensive analysis of impact of the parameter $\beta$ on the behavior of \fddscore. We contend that the flexibility  offered by the  parameter  $\beta$ represents an important advantage over other existing term-weighting schemes. 
    
    The second contribution is an extensive comparative evaluation between the \fddscore weighting scheme and eighteen weighting schemes. The schemes used for comparison are based on traditional and state of the art methods, including both unsupervised and supervised approaches. The code for the \fddscore score and the eighteen replicated schemes are made available to the research community for reproducibility.  The comparative performance evaluation showed that despite its simplicity, \fddscore achieves results that are competitive with state-of-the-art methods that rely on more complex information-theoretic and statistical notions. 
    
    The third contribution is new insight into  the question of how to build multi-term queries based on supervised term-weighting strategies. The analysis presented in this article demonstrated that the \fddscore offers a useful mechanism to explore different approaches to build complex queries.
  
  Finally, the fourth contribution is a publicly available data set labeled by experts in Economy that can be used to compute topic-dependent term weights or applied to other supervised task in the economic domain. In this work, the data set was used to carry out a comprehensive analysis of the behavior of  \fddscore  as a function of the adjustable parameter $\beta$.

     As part of our future work we  plan to define other retrieval strategies based on \fddscore that introduce the formulation of queries with more complex syntaxes. The \fddscore weighting scheme will  also be evaluated on classification tasks, which will open new challenges. In particular, we will investigate techniques to learn optimal $\beta$ values for term-weighting during training.  We will also look into the problem of adapting  \fddscore (which relies on class labels available during training) to derive a weighting scheme for terms in the test and validation sets, where the class labels are not available.   
   
\section*{Acknowledgment}
    This work was supported by   CONICET, Universidad Nacional del Sur (PGI-UNS 24/N051 and PGI-UNS 24/E145) and by a LARA project (Google Research Award for Latin America 2019).

\newpage

  \bibliographystyle{apalike}

\end{document}